# Simple fault-tolerant encoding over q-ary CSS quantum codes

Pedro J. Salas[*]

Depto. Tecnologías Especiales Aplicadas a la Telecomunicación, E.T.S.I. Telecomunicación, U.P.M., Ciudad Universitaria s/n, 28040 Madrid (Spain)

Condensed title: Fault-tolerant encoding

## ABSTRACT

CSS codes are a subfamily of stabilizer codes especially appropriate for fault-tolerant quantum computations. A very simple method is proposed to encode a general qudit when a Calderbank-Shor-Steane quantum code, defined over a q-ary alphabet, is used.

## 1. INTRODUCTION

Quantum computation using a small number of qudits (or qubits if the dimension of the Hilbert space is two) is, now, a reality. When the required precision in the computation is not a crucial factor, decoherence can be overcome isolating the devices and carefully manipulating the systems. Unfortunately, if our objective is to carry out long time quantum computations, a good active and/or passive stabilization method has to be used to avoid decoherence[1]. One of these active methods is quantum error correcting codes (QECC) that protect the qudit information spreading it in quantum registers by means of a general process called encoding.

Once the information is encoded, the quantum algorithm is carried out applying the gates directly onto the encoded qudits and, at the end of the process; the results are recovered decoding the information. At some specified time steps during the computation, error correction procedures are carried out in order to circumvent the decoherence. In this situation a new problem arises because, the error correction step is itself, noisy. Therefore, we have to be very careful to design correction networks that do not introduce more errors than they try to eliminate. This is the main idea underlying the fault-tolerant quantum computing methods[2,3,4]. Several ingredients are necessary to achieve fault-tolerance: encoding through a QECC, encoded logic not spreading the errors perniciously, an appropriate and periodic error correction step and, not very complicated networks involved in each of these steps.

The Calderbank-Shor-Steane (CSS) codes[5,6] were the first family of QECC discovered. It is currently known they are a subfamily of the stabiliser codes, having some special and simplifying properties. One of these is the transversality of the encoded CNOT gate[7], guaranteeing its fault-tolerance. A CNOT gate is transversal between two encoded qudits, if the encoded gate is implemented as a sequence of CNOT gates connecting different control qudits with different target qudits. This encoded CNOT gate for CSS codes is very simple and will simplify all the networks in which it is involved.

In this paper we consider a very simple and economical method to encode a qudit (or qubit if the dimension is 2) when the code is a CSS over a q-ary alphabet. The main point is to synthesize fault-tolerantly an encoded zero $|0_E\rangle$, after that, any other basis state or general qudit can be encoded straightforwardly. Firstly, the q = 2 (binary) case will be considered (CSS codes over the field $\mathbb{F}_2$), introducing the detailed networks. Next, the treatment will be generalized to qudits in case of CSS codes over $\mathbb{F}_q$. We assume the independent error model; even though the QECC can be used if some correlations are still present in the system. When a QECC correcting t errors is used to encode the information, a quantum circuit is considered fault-tolerant if the probability of uncorrectable errors behaves as $O(\varepsilon^{t+1})$, $\varepsilon$ being the error probability per qudit and time step.

---

[*] Electronic mail psalas@etsit.upm.es

## 2. BINARY CSS CODES

The simplest way to represent quantum information is by means of a system with the Hilbert space $\mathbb{C}^2$ of dimension two (q=2) called qubit, $\{|0>, |1>\}$ being the orthonormal basis. From the point of view of quantum codes the information can be encoded as quantum registers $|q_1 q_2 \ldots q_n> \in \mathbb{C}^{2^n}$, built with strings of symbols of the binary alphabet $q_i \in \{0,1\} = \mathbb{F}_2$.

The (real) Pauli group for one qubit is defined as $P_1 = \pm\{I, X=\sigma_X, Y=-i\sigma_Y, Z=\sigma_Z\}$, I being the identity and $\sigma_{X,Y,Z}$ the Pauli matrices. In a similar way the n-qubit Pauli group is defined as $P_n = \{\pm X_r Z_s, r,s \in \mathbb{F}_2^n\}$. The notation used for the operator $X_r$ (and $Z_s$) means the application of an operator X (respectively Z) to the i-th qubit if the coordinate $r_i$ of the vector $r \in \mathbb{F}_2^n$ is $r_i = 1$, and the identity if $r_i = 0$. A binary stabilizer quantum code $Q = [[n,K,d]]_2$ is defined as a non-empty subspace of $\mathbb{C}^{2^n}$ built as the common eigenspace of an abelian subgroup S (called stabilizer) of the Pauli group $P_n$:

$$Q = \bigcap_{\substack{E \in S \\ S \leq P_n}} \left\{ |q\rangle \in \mathbb{C}^{2^n}, E|q\rangle = |q\rangle \right\} \tag{1}$$

CSS codes are a subfamily of the stabilizer quantum codes, having a concrete structure in the set of generators (called $\langle S \rangle$) of the S subgroup: they can be written not showing Y factors. Their construction can be made starting from two classic codes $C_1 = [n,k_1,d_1]$ with parity check matrix is $H_1[(n-k_1)\times n]$ and $C_2 = [n, k_2,d_2]$ with parity check matrix $H_2[(n-k_2)\times n]$, fulfilling the condition $C_2^\perp \leq C_1$ ($C_2^\perp$ is the usual Euclidean orthogonal code). The subcode $C_2^\perp$ induces a partition on $C_1$ in $2^{k_1+k_2-n}$ cosets defining the code states and a quantum CSS code with parameters $Q_{CSS} = [[n, K=k_1+k_2-n, d\geq\min(d_1,d_2)]]_2$ that is pure to $\min(d_1,d_2)$. A quantum code is called pure if distinct elements of Pauli group produce orthogonal states. A slightly less general construction assumes $C_1 = C$ and $C_2^\perp = C^\perp$, and the starting point are codes fulfilling the condition $C^\perp \leq C$ ($C^\perp$ is a weakly self-dual code) and the quantum codes have the parameters $Q_{CSS} = [[n,K=2k-n,d]]_2$. Throughout this paper we will only consider CSS codes satisfying this weakly self-duality condition, because the assumption is not too restrictive and simplifies the notation. The states of the $Q_{CSS}$ code have the following aspect:

$$|u_E\rangle = \frac{1}{\sqrt{|C^\perp|}} \sum_{x \in C^\perp} |x \oplus uD\rangle \tag{2}$$

D being a (K×n) matrix of coset leaders (K vectors of length n and minimum weight one per coset) and u ∈ $\mathbb{F}_2^K$ (the addition is modulo 2).

The fault-tolerant encoding problem using CSS codes is concentrated mainly on synthesizing $|0_E\rangle$ states. The process will take several steps. We distinguish two cases.

### 2.1 Non-fault-tolerant synthesis of the $|0_E\rangle$ state

The particular structure of the CSS quantum codes permits an easy implementation of the encoding network[8,9]. Given a CSS binary code $[[n,K,d]]_2$ built by means of the classic codes $C^\perp = [n,n-k,d^\perp]_2 \subseteq C = [n,k,d']_2$, the encoding operation is an application E: $\mathbb{C}^k \to \mathbb{C}^n$ from the Hilbert subspace of dimension k to a Hilbert space of greater dimension n. According the equation (2), the encoded zero state ($|0_E\rangle$) is the superposition of all $C^\perp$ codewords. It is not difficult to construct a quantum network implementing such an encoding if we realize the structure of the codewords. The generation matrix of $C^\perp$ in the standard form has the structure $G_{C^\perp} = (I_{n-k} | A_{n-k \times k})$, $I_{n-k}$ being the identity of dimension (n-k), corresponds to the information qubits and A is a matrix containing the encoding information. Each $C^\perp$ codeword includes a piece of information qubits having all possible binary strings of length (n-k) and the remaining k qubits contain additional information according to the rows in A. The information qubit string can be created via (n-k) Hadamard gates applied to the (n-k) first qubits. The flipping process to produce the A section is carried out using CNOT gates with the control qubit on each information qubit and the target situated on the appropriate qubit among the last k qubits.

Obviously, this $|0_E\rangle$ synthesis is not fault-tolerant because several CNOT gates connect the same control qubit with different target qubits inside the same quantum codeword. Therefore a bit-flip error in the control qubit spreads to several control qubits.

### 2.2 Fault-tolerant synthesis of the $|0_E\rangle$ state. Case of C being a perfect code

Consider a $Q_{CSS} = [[n,K,d]]_2$ constructed from the classic code $C = [n,k,d']_2$. The $|0_E\rangle$ state (0 ∈ $\mathbb{F}_2^K$) is the linear combination of all the $C^\perp$ codewords, and the quantum network accomplishing the task involves only Hadamard (H) and CNOT gates. The H gates are applied directly to the $|0\rangle$ qubits and do not spread the possible errors appearing in them, so its application is fault-tolerant. CNOT gates are introduced according the generation matrix of $C^\perp$. As is well known, CNOT gates propagate bit-flip errors (X) forward, meanwhile the phase errors (Z) are propagated backward[2,3,4]. As a consequence, CNOT gates connecting the same target qubit with different control qubits (and vice versa), allow error spreading within the quantum register and destroy the fault-tolerance of the network.

However, some CSS codes are intrinsically robust to phase-flip errors and $Z_v$ (with W(v) > t = $\lfloor(d-1)/2\rfloor$, W(·) being the usual Hamming weight) never appear in the $|0_E\rangle$ states. Therefore, in synthesizing $|0_E\rangle$ state we shall not worry about this type of error. The following theorems verify it.

***Theorem 1***
Let $Q_{CSS} = [[n,K,d]]_2$ be obtained from $C = [n,k,d']_2$, fulfilling $C^\perp \leq C$. Each code state of $Q_{CSS}$ is the linear combination of the quantum registers in each coset $C_u$:

$$C_u = \left\{ |x \oplus uD\rangle, x \in C^\perp \right\}, u \in \mathbb{F}_2^K \qquad (3)$$

D(K×n) being the matrix of coset leaders. The condition

$$\bigcup_{\substack{a \in \mathbb{F}_2^n, W(a) \leq t \\ u \in \mathbb{F}_2^K}} X_a C_u = \mathbb{F}_2^n \qquad (4)$$

is fulfilled if and only if C is a perfect classic code.

□ This theorem establishes the condition under which the $X_a$ translations until weight t of all the registers involved in $Q_{CSS}$ produces $\mathbb{F}_2^n$. If the number of $Q_{CSS}$ states is $2^K$, the number of registers per $Q_{CSS}$ state is $2^{n-k}$ and assuming the condition of the theorem is fulfilled, then:

$$2^K \times 2^{n-k} \times \sum_{i=0}^{t} \binom{n}{i} = 2^n \qquad (5)$$

If K = 2k-n, the Hamming bound is reached for C. Inverting the process we obtain the sufficient condition. ∎

*Corollary*
Consider a $Q_{CSS} = [[n,K,d]]_2$ coming from a C perfect classic code. Given an operator $X_v$, $v \in \mathbb{F}_2^n$ with $W(v) > t = \lfloor (d-1)/2 \rfloor$, then there is a v' $\in \mathbb{F}_2^n$, $W(v') \leq t$ such as:

$$X_v \sum_{u \in \mathbb{F}_2^K} |u_E\rangle = X_{v'} \sum_{u \in \mathbb{F}_2^K} |u_E\rangle \qquad (6)$$

□ As a consequence of the fact that $X_v$ with $W(v) > t$ does not create new quantum registers when applied to all $Q_{CSS}$ states, and $X_v$ does not change the relative phases of the linear combination. ∎

We can now demonstrate the theorem establishing the non-appearance of phase-flip errors of weight greater than t.

*Theorem 2*
Consider a $Q_{CSS} = [[n,K,d]]_2$ coming from a C perfect classic code, then the $|0_E\rangle$ (0 $\in \mathbb{F}_2^K$) state cannot have phase-flip errors $Z_v$, $v \in \mathbb{F}_2^n$, with $W(v) > t = \lfloor (d-1)/2 \rfloor$.

□ Let $Z_v$ be an error with $W(v) > t$, $v \in \mathbb{F}_2^n$,

$$Z_v |0_E\rangle = H_{trans} X_v H_{trans} |0_E\rangle = H_{trans} X_v H_{trans} \sum_{x \in C^\perp} |x\rangle = H_{trans} X_v \sum_{y \in C} |y\rangle$$

$H_{trans}$ is a transversal Hadamard gate and the last step of the previous equation comes from the dual code theorem. As $C^\perp \leq C$, $C^\perp$ and D generate C, then $\forall y \in C$, $y = x \oplus uD$ with $x \in C^\perp$ and $u \in \mathbb{F}_2^K$. Then

$$\sum_{y \in C} |y\rangle = \sum_{\substack{x \in C^\perp \\ u \in \mathbb{F}_2^K}} |x \oplus uD\rangle = \sum_{u \in \mathbb{F}_2^K} |u_E\rangle$$

The constants are not included. Therefore

$$Z_v |0_E\rangle = H_{trans} X_v \sum_{u \in \mathbb{F}_2^K} |u_E\rangle = H_{trans} X_{v'} \sum_{u \in \mathbb{F}_2^K} |u_E\rangle = H_{trans} X_{v'} H_{trans} |0_E\rangle = Z_{v'} |0_E\rangle$$

Where in the second step the earlier corollary has been taken into account, establishing the existence of a v' $\in \mathbb{F}_2^n$, W(v') $\leq$ t fulfilling the condition. So we conclude that $Z_v \equiv Z_{v'}$ for |0$_E$> with W(v') $\leq$ t. ∎

According to the previous theorems, the Q$_{CSS}$ built from binary classic perfect codes involve |0$_E$> states such as they are robust to phase-error accumulation. This fact will simplify the networks for the synthesis of |0$_E$> states. We will look for classic binary codes which are perfect and weakly self-dual. The only perfect (linear and non trivial) binary codes are the Golay G$_{23}$ and Hamming codes[10]. The binary Golay code [23,12,7]$_2$ has an even subcode [23,11,8]$_2$ through which the CSS [[23,1,7]]$_2$ quantum code can be obtained, correcting errors of weight three. The Hamming codes have the parameters $C_H$ = [$2^m$-1, $2^m$-1-m,3]$_2$ and their Euclidean duals are the simplex codes $C_H^\perp$ = [$2^m$-1, m,$2^{m-1}$]$_2$, all of them having m(whole) $\geq$ 2. Simplex codes $C_H^\perp$ are weakly self-duals for m $\geq$ 3 and can be used to construct Q$_{CSS}$ = [[$2^m$-1, $2^m$-2m-1,3]]$_2$. The simplest case is the well known Steane's code [[7,1,3]]$_2$. Some results were found with this code in[11].

In the synthesis of a |0$_E$> state for those CSS codes, phase-flip errors will never have a weight greater than t = $\lfloor$(d-1)/2$\rfloor$. The |0$_E$> state will be prepared according to the generation matrix ($G_{C^\perp}(n-k \times n)$) of C$^\perp$: applying (n-k) Hadamard gates and the adequate CNOT gates with the control on the (n-k) previous qubits. The only task to be carried out is to detect the bit-flip errors of weight greater than t. To explain the process we will use the explicit notation for the errors before the state when their weight can be greater than t and as a subindex (with an over bar) when its weight is less than t. The first step is to synthesize (t+1) encoded zero states starting from |0$^{\otimes n}$>, according to $G_{C^\perp}(n-k \times n)$. The states obtained will be called $X_{u_i} |0_E\rangle_{\overline{Z}}$ (i=0,...,t), because they have some bit-flip errors but not phase-flips. Although the states do not include phase errors with weight greater than t, the synthesis is not fault-tolerant because bit-flip errors $X_{u_i}$ with W(u$_i$) $\leq$ 3 still appear with probability O($\epsilon$), $\epsilon$ being the error probability for time step and qubit. To eliminate the bit-flip errors, the network of figure 1 is used. For the CSS codes coming from Hamming classic codes with d = 3, the general network of figure 1 will include only two encoded zero states: $X_{u_0} |0_E\rangle_{\overline{Z}}$ and $X_{u_1} |0_E\rangle_{\overline{Z}}$. The method proposed in this paper to prepare zero states fault-tolerantly, implies taking advantage of the fact that transversal (and then fault-tolerant) application of a CNOT gate implements an encoded CNOT gate on every CSS code[7]. An encoded CNOT gate between the states $X_{u_0} |0_E\rangle_{\overline{Z}}$ and $X_{u_1} |0_E\rangle_{\overline{Z}}$ transfers forward the $X_{u_0}$ to

$X_{u_1}|0_E\rangle_{\bar{Z}}$ to get the state $X_{u_0 \oplus u_1}|0_E\rangle_{\bar{Z}}$. Next, this second state is destructively measured and a classic register $w \in \mathbb{F}_2^n$ is obtained. Unless $w \in C^\perp$ and fulfilling $H_{C^\perp} w^T = 0$, the state $X_{u_0}|0_E\rangle_{\bar{Z}}$ is rejected and the process is restarted with another pair of synthesized zero states. The error probability comes from the fact that $u_0 \oplus u_1 = 0$, but this happens with a probability $O(\varepsilon^2)$ and the method is fault-tolerant. When the process is successful, we will call the final zero state achieved $|0_E\rangle_{\overline{XZ}}$ to indicate that it has no phase or bit-flip errors except those allowed by the fault-tolerant method. Note that if the second state has no bit-flips (and $u_1 = 0$), the code $C^\perp$ is able to detect *any* bit-flip error on the first zero state.

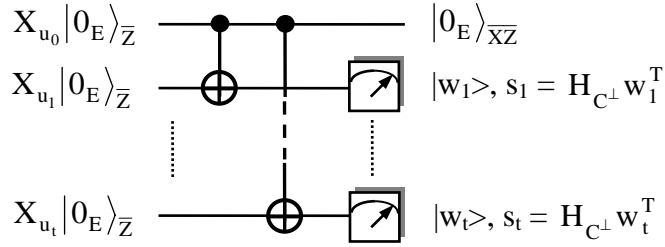

FIG. 1. Network to eliminate bit-flip errors from the first $X_{u_0}|0_E\rangle$ state previously synthesized, by CNOT the state with t- $X_{u_i}|0_E\rangle_{\bar{Z}}$ states. After a destructive measurement, the classic register allows us to calculate the syndrome by means of the parity check matrix of $C\perp$. If all the syndromes $s_i = 0$, in the first state ($|0_E\rangle_{\overline{XZ}}$) the bit-flips with weight greater than t, appear with probability $O(\varepsilon^{t+1})$.

If the CSS coming from the Golay code is used (d = 7), to eliminate the bit-flips fault-tolerantly we will need four $X_{u_i}|0_E\rangle_{\bar{Z}}$ states in figure 1. The first zero state will be accepted if and only if the syndromes $s_i$ (i=1, 2, 3) are zero, otherwise the process will be restarted with four new states. The error probability, i.e. the probability of having a bit-flip error $X_v$ with $W(v) \geq 4$ in the first zero state, behaves as $O(\varepsilon^4)$, then the process is fault-tolerant.

2.3 Fault-tolerant synthesis of the $|0_E\rangle$ state. Case of C not being a perfect code

When the C code is not perfect, the CSS code is not restricted to those obtained from Golay or Hamming classic codes. The $Q_{CSS} = [[n,K,d]]_2$ could have phase-flip errors of weight greater than t and both of them must be taken into account in the zero state synthesis. One phase-flip error can be spread backward by CNOT gates making the networks non fault-tolerant to those errors.

In this case the first step is to synthesize (t+1) zero encoded states according the $C^\perp$ generation matrix to obtain the states $X_{u_i} Z_{v_i}|0_E\rangle$ (i=0,...,t) affected by bit and phase-flip errors of any weight. The phase-flips are then eliminated fault-tolerantly using the general network detailed in figure 2. After the second Hadamard, the states are measured destructively to find out the syndromes $\{s_i, i=1,..,t\}$. If $s_i = 0$ (i=1,..,t), the first state is considered $O(\varepsilon^{t+1})$ free of phase-flips, and the achieved state is $X_{u_0}|0_E\rangle_{\bar{Z}}$. Each piece of the network $H_i$ CNOT(0,i) $H_i$ connecting the first state zero with the i-th, transfers the $Z_{v_0}$ error backward to $X_{u_i} Z_{v_i} |0_E\rangle$. The process is shown explicitly to clarify its fault-tolerance:

$$X_{u_i} Z_{v_i} |0_E\rangle \xrightarrow{H_i} Z_{u_i} X_{v_i} H |0_E\rangle \xrightarrow{CNOT(0,i)} Z_{u_i \oplus v_0} X_{v_i} H |0_E\rangle$$
$$= (-1)^{(u_i \oplus v_0) \cdot v_i} X_{v_i} Z_{u_i \oplus v_0} H |0_E\rangle \xrightarrow{H_i} (-1)^{(u_i \oplus v_0) \cdot v_i} Z_{v_i} X_{u_i \oplus v_0} |0_E\rangle \quad (7)$$

In the sequence we have considered that H H = I, H $Z_v$ H = $X_v$ and $Z_u X_v$ = $(-1)^{u \cdot v} X_v Z_u$. The bit-flip error $X_{u_i \oplus v_0}$ can be detected by means of the parity check matrix of $C^\perp$, unless $u_i \oplus v_0 = 0$, but this happens with probability $O(\varepsilon^2)$ for the i-th syndrome, if the errors are independent. The whole process would fail if all of the t syndromes $s_i = 0$ and $v_0 \neq 0$; but this event has a probability $O(\varepsilon^{t+1})$.

Once we have a set of states { $X_{u_i} |0_E\rangle_{\bar{Z}}$, i=0,1,..,t}, the network of figure 1 can be used to eliminate the bit-flip errors from the state i=0, to finally obtain the $|0_E\rangle_{\overline{XZ}}$ state. Note that the inverted process would not be advantageous. If the first step was the bit-flip elimination by means of the network in figure 1, the obtained states would be $Z_{v_i} |0_E\rangle_{\bar{X}}$. If these states are then used in the network of figure 2 to eliminate phase-flips, the first Hadamard rotation would transform $Z_{v_i}$ into $X_{v_i}$ in the dual basis and then spread forward to the $Z_{v_0} |0_E\rangle_{\bar{X}}$ state, reintroducing bit-flips on it again.

This filtering process reminds us of the method used by Steane in[12]. In that case, the ancilla state $|0_E\rangle$ is checked only for one type of error; let us say bit-flips ($X_v$), by means of a set of control-Z gates from the verification state $|+\rangle$ to the ancilla state. The verification checks are carried out according the observables in the stabilizer of $|0_E\rangle$. Finally, the number of time steps ($N_t$) needed to carry out the process is optimized by means of a strategy based on the Latin rectangle. The method provides $N_t$ = N+1 (the final step previous to the measurement, is not included), N being the largest weight of a row or column of the A matrix involved in the standard form of the parity check matrix $H_{C^\perp} = \left( A^T_{k \times n-k} | I_k \right) = G_C$ ($I_k$ being the identity matrix) of the classic code $C^\perp$ = [n,n-k,$d^\perp$] providing the $|0_E\rangle$ state. Taking into account this construction, the number of time steps that are to be carried out is $N_t = d^\perp$, while checking a noisy $X_u Z_v |0_E\rangle$ state for bit-flip errors using the present scheme (see figure 1), only involves *two time steps*, assuming we have t states $X_{u_i} Z_{v_i} |0_E\rangle$ (i=1,..,t) already synthesized. Moreover, the following lemma establishes that the present method is less costly with regard to the number of two-qubit gates used when the $Q_{CSS}$ is a pure code. It is already known[15] that given two classic codes fulfilling $C^\perp$ = [n,n-k,$d^\perp$] ≤ C = [n,k,d']$_2$ there exists a CSS code [[n,K,d]]$_2$ being pure to d = min($d^\perp$, d'), and d = d' because $C^\perp$ ≤ C.

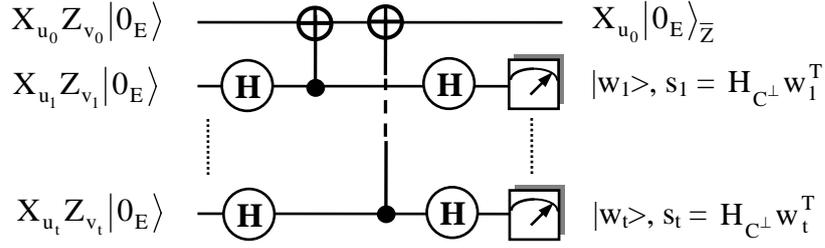

FIG. 2. Network to eliminate phase-flip errors from the first $X_{u_0}Z_{v_0}|0_E\rangle$ state previously synthesized, by CNOT the state with t- $X_{u_i}Z_{v_i}|0_E\rangle$ states. After a destructive measurement, the classic register allows us to calculate the syndrome by means of the parity check matrix of $C^\perp$. If all the syndromes $s_i = 0$, in the first state ($X_{u_0}|0_E\rangle_{\overline{Z}}$) the phase-flips with weight greater than t, appear with probability $O(\varepsilon^{t+1})$.

## *Lemma 1*

Let $Q_{CSS} = [[n,K,d]]_2$ be obtained from $C = [n,k,d']_2$, fulfilling $C^\perp \leq C$ and being pure to $d = \min(d^\perp, d') = d'$. The number of control-Z gates[12] $g_Z \geq g_X$ (number of CNOT gates used in the present method).

□ The number of control-Z gates ($g_Z$) is the number of "1" in the $H_{C^\perp}$ matrix, which means $g_Z \geq kd'$, because d' is the minimum weight of row in the parity check matrix. Because n-k ≤ k then k ≥ n/2 and:

$$g_Z \geq kd' \geq (n/2)\, d' = (n/2)\, d \geq nt = g_X$$

where we have used $d \geq 2t+1$. The number $g_X$ comes from using t $X_{u_i}Z_{v_i}|0_E\rangle$ (i=1,..,t) states connected transversally with n CNOT gates. ∎

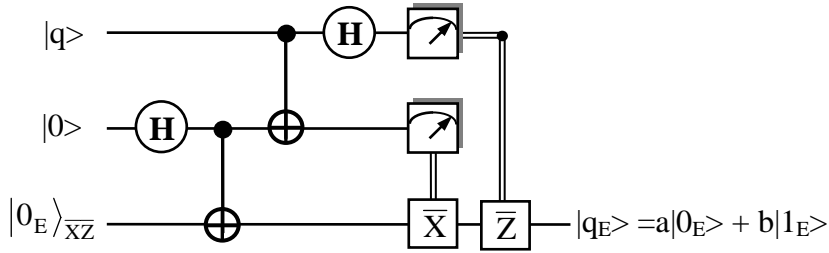

FIG. 3. Network to encode a qubit |q> = a|0> + b|1> fault-tolerantly. The |q_E> is the encoded qubit $|0_E\rangle_{\overline{XZ}}$ being the encoded zero state of the CSS code and $\overline{X}$ and $\overline{Z}$ are encoded Pauli gates. Note that the first CNOT gate involves an encoded NOT gate. The |q_E> state is also affected by some bit and phase-flip errors, but always of weight less than t.

Then, the method considered in this paper seems to be more economic in gate and time steps, so it will introduce fewer errors in the |0_E> synthesis.

## 2.4 Encoding a general qubit

Once a fault-tolerant $|0_E\rangle_{\overline{XZ}}$, $0 \in \mathbb{F}_2^K$ for $Q_{CSS} = [[n,K,d]]$ has been synthesized, the code basis can be easily generated applying the encoded Pauli operators of the normalizer transversally (and then fault-tolerantly).

On the other hand, taking advantage of the transversal implementation of the Hadamard and CNOT gates, as well as the encoded Pauli operators, we can use the teleportation circuit[13,14] shown in figure 3 to encode any qubit fault-tolerantly. The circuit is fault-tolerant because only needs zero encoded states and gates being transversal in CSS codes.

## 3. q-ARY CSS CODES

The binary alphabet could sometimes be too restrictive. Leaving the restriction means defining quantum codes over $\mathbb{F}_q$ fields. In this case the quantum information is represented by means of a quantum system with Hilbert space $\mathbb{C}^q$ called qudit. From the point of view of the code, the information can be represented as quantum registers $|q_1 q_2 ... q_n\rangle \in \mathbb{C}^{q^n}$, built with strings of symbols of the q-ary alphabet $q_i \in \mathbb{F}_q$. To construct the q-QECC we generalize those of binary codes.

Suppose $q = p^m$ (p prime) and denoted by $\omega$ a primitive complex p-th root of unity ($\omega = \exp(2\pi i/p)$), the Pauli group for one qudit is defined as $P_1 = \{\omega^a X_r Z_s, r,s \in \mathbb{F}_q, a \in \mathbb{F}_p\}$. The operators $X_r$ and $Z_s$ are defined as:

$$X_r |x\rangle = |x+r\rangle$$
$$Z_s |x\rangle = \omega^{tr(s \cdot x)} |x\rangle \quad x,r,s \in \mathbb{F}_q \quad (8)$$

The $tr(\cdot)$ means the trace of an element $x \in \mathbb{F}_q$, $tr(x) = \sum_{i=0}^{m-1} x^{p^i}$. The n-qudit Pauli group is defined as the tensor product of n factors of $P_1$. A q-ary stabilizer quantum code[15] $Q = [[n,K,d]]_q$ is defined as a non-empty subspace of $\mathbb{C}^{q^n}$ that is common of an abelian subgroup S (called stabilizer) of the Pauli group $P_n$:

$$Q = \bigcap_{\substack{E \in S \\ S \leq P_n}} \{|q\rangle \in \mathbb{C}^{q^n}, E|q\rangle = |q\rangle\} \quad (9)$$

The q-ary CSS codes are a subfamily of the stabilizer quantum codes, and can be constructed in a similar way as the binary codes. Starting from two classic codes $C_1 = [n,k_1,d_1]_q$ and $C_2 = [n,k_2,d_2]_q$, fulfilling the condition $C_2^\perp \leq C_1$ ($C_2^\perp$ is the usual Euclidean orthogonal code) then there is a quantum CSS code with the parameters $Q_{CSS} = [[n, K=k_1+k_2-n, d = \min(W(c), c \in (C_1 \backslash C_2^\perp) \cup (C_2 \backslash C_1^\perp))]]_q$. Analogous to the binary case, a slightly less general construction assumes $C_1 = C$ and $C_2^\perp = C^\perp$, and the starting point are codes fulfilling the condition $C^\perp \leq C$ ($C^\perp$ is a self-orthogonal code). The quantum codes have the parameters $Q_{CSS} = [[n,K=2k-n,d]]_q$. The states of the $Q_{CSS}$ code have the same aspect as was shown in equation (2) with $D(K \times n)$ being the matrix of coset leaders and $u \in \mathbb{F}_q^K$.

We can follow the tracks of what was done in the binary case, to encode using these CSS codes. The first step is to synthesize the zero state according the generation matrix of $C^\perp$, having the following aspect $G_{C^\perp} = (I_{n-k}|A_{n-k\times k}) \in \mathbb{F}_q$. The encoding involves single qudit gates as Fourier gates F and multiplying gates $M_r$, and the ADD two qudit gate (q-version of a CNOT), defined as

$$F = \frac{1}{\sqrt{q}} \sum_{x,y \in \mathbb{F}_q} \omega^{tr(x\cdot y)} |y\rangle\langle x| \qquad M_r = \sum_{x \in \mathbb{F}_q} |rx\rangle\langle x| \quad \text{for } r \in \mathbb{F}_q - \{0\} \qquad (11)$$

$$ADD(1,2) = \sum_{x,y \in \mathbb{F}_q} |x\rangle\langle x|_1 \otimes |x+y\rangle\langle y|_2 \qquad (12)$$

The encoding network[16] involves (n-k) F gates to create the initial linear combination of those first information qudits. The construction $M_r(2)$ ADD(1,2) $M_{r^{-1}}(2) = \sum_{x,y} |x\rangle\langle x| \otimes |rx+y\rangle\langle y|$ add rx to the second qudit and allow us to build the piece of the circuit corresponding to the A section of the $G_{C^\perp}$ matrix, controlled by the $I_{n-k}$.

The F and $M_r$ gates can be implemented transversally in the encoding networks and they are fault-tolerant. For these CSS codes the transversal ADD gate implements the encoded version and the main problem is raised by its error spreading. As is well known[17] the bit and phase-flips are propagated by the ADD gate as: $X_r \otimes I \to X_r \otimes X_r$, $I \otimes X_r \to I \otimes X_r$, $Z_r \otimes I \to Z_r \otimes I$, $I \otimes Z_r \to Z_r^{-1} \otimes Z_r$. Note the backward propagation of phase-flips as their inverse.

It is not difficult transform theorem 1, corollary and theorem 2 to the present q-ary CSS codes, as well as their consequences. We only sketch

*Theorem 3*
Consider a $Q_{CSS} = [[n,K,d]]_q$ coming from a C perfect classic q-ary code, then the $|0_E\rangle$ ($0 \in \mathbb{F}_q^K$) state cannot have phase-flip errors $Z_v$, $v \in \mathbb{F}_q^n$, with $W(v) > t = \lfloor(d-1)/2\rfloor$.

□ Let $Z_v$ be an error with $W(v) > t$, $v \in \mathbb{F}_q^n$,

$$Z_v|0_E\rangle = F_{trans} X_v F_{trans}^+ \left\{\sum_{x \in C^\perp} |x\rangle\right\} = F_{trans} X_v \sum_{y \in C} |y\rangle = F_{trans} X_v \sum_{u \in \mathbb{F}_q^K} |u_E\rangle =$$

$$= F_{trans} X_{v'} \sum_{u \in \mathbb{F}_q^K} |u_E\rangle = F_{trans} X_{v'} F_{trans}^+ |0_E\rangle = Z_{v'}|0_E\rangle$$

No constants are considered in the states. In the second step we have taken into account the condition $\sum_{x \in C^\perp} \omega^{-tr(y \cdot x)} = \begin{cases} q^{n-k} & \text{if } y \in C \\ 0 & \text{if } y \notin C \end{cases}$ ∎

Therefore we will look for q-ary perfect and self-orthogonal classic codes. The only perfect (linear and non-trivial) q-ary codes are[10] the ternary (q=3) Golay $G_{11}$ = $[11,6,5]_3$ and the Hamming codes. The dual of the Golay code $[11,5,6]_3$ is self-orthogonal and can be used to obtain the $Q_{CSS} = [[11,1,5]]_3$, correcting two errors. The q-ary Hamming codes have the parameters $C_H = [n, n-m,3]_q$ with $n = q^m-1/q-1$ and their Euclidean duals are the simplex codes $C_H^\perp = [n, m,q^{m-1}]_q$ ($m \geq 2$). Simplex codes $C_H^\perp$ are self-orthogonal and can be used to construct $Q_{CSS} = [[n, n-2m,3]]_q$, however they are not optimal because they do not meet the quantum Hamming bound[15]. The symmetry of the previous codes avoids phase-flip errors accumulating dangerously, and then we only have to worry about bit-flip errors. To synthesize zero states fault-tolerantly, we will use the network of figure 1 with the same strategy as was used for the binary codes. All the syndromes $\{s_i, i =1,..,t\}$ must be zero, otherwise the network will be restated with (t+1) new states.

When the classic codes C are not perfect, they will be affected by phase and bit-flip errors of any weight. Both of them are spread via ADD gates. Following the same strategy as for binary codes, we first detect the phase-flips by means of the network of figure 2 and then the bit-flips with the network of figure 1, to avoid the dangerous bit-flip spreading.

The encoding of a general qudit can be carried out through the teleportation network shown in figure 3. Again, all the involved gates can be implemented transversally on the CSS codes over q-ary alphabets, therefore being fault-tolerant.

## 4. CONCLUSIONS

A very simple method to encode states fault-tolerantly by means of q-ary CSS quantum codes is provided. We demonstrate that in case of a CSS $[[n,K,d]]_q$ code, constructed from a classic perfect code C, the quantum code has no phase-flips of Hamming weight greater than $\lfloor (d-1)/2 \rfloor$, then the fault-tolerance is only concerned with a bit-flip error control. Once a zero state has been synthesized, the encoded gates of the normalizer are able to generate the encoded basis. In addition, a general qudit can be encoded by means of a teleporting circuit that for CSS codes involves transversal fault-tolerant gates.

## ACKNOWLEDGMENTS

The author would like to thank the financial support from the Spanish Project CCG06-UPM/INF-389.